\documentstyle[mprocl]{article}

\def\apj#1{{\em Astrophys. J.} {\bf #1}}

\def\aa#1{{\em Astron. Astrophys.} {\bf #1}}

\def\nat#1{{\em Nature} {\bf #1}}
\def\apjs#1{{\em Astrophys. J. Suppl.} {\bf #1}}

\def\be{\begin{equation}}
\def\ee{\end{equation}}
\def\bea{\begin{eqnarray}}
\def\eea{\end{eqnarray}}
\def\etal{{\it et al.}\ }
\def\Mpc{$h^{-1}$~{\rm Mpc}}

\begin{document}

\title{HAS THE UNIVERSE A HONEYCOMB STRUCTURE?}

\author{J. EINASTO}

\address{Tartu Observatory, EE-2444 T\~oravere, Estonia}

\maketitle

\abstracts{Recent analysis of the distribution of clusters of galaxies
is reviewed. Clusters of galaxies located in rich superclusters form a
quasiregular lattice similar in structure to honeycombs. The power
spectrum of clusters of galaxies has a sharp peak at wavelength
$\lambda_0=120$~\Mpc\ corresponding to the lattice step.  The peak in
the spectrum may be due to processes during the inflationary stage of
the structure evolution.  }
  
\section{Introduction}

According to the current paradigm of the large-scale structure of the
Universe  galaxies and clusters of galaxies are concentrated to elongated
filamentary chains and the space between filaments is void of galaxies.
Such distribution resembles cells \cite{zes82};   examples are the
Northern Local Void surrounded by the Local, Coma, and Hercules
superclusters,  and the Bootes void located between the Hercules and
Bootes supercluster \cite{l95}. 

Superclusters and voids form a continuous network of alternating high-
and low-density regions \cite{me}.  They are formed by density waves of
wavelength corresponding to the scale of the supercluster-void
network. It is believed that density waves have a Gaussian
distribution, thus high- and low-density regions should be randomly
distributed.  It was therefore a great surprise when Broadhurst
\etal \cite{beks90} found that the distribution of high-density
regions in a small area around the northern and southern Galactic pole
is fairly regular: high- and low-density alternate with rather
constant step of 128~\Mpc.  The regularity is so far well established
only in the direction of Galactic polar caps, in other directions the
regularity is much less pronounced. In order to find the degree of
regularity of the supercluster-void network 3-dimensional data of the
distribution of high-density regions are needed.  For this purpose
Abell-ACO clusters of galaxies \cite{abell} can be used, as they form
the largest and deepest sample of astronomical objects covering the
whole celestial sphere outside the Milky Way zone of avoidance.

\begin{figure*}[t]
\vspace*{6cm}
\caption{Distribution of clusters in high-density regions in supergalactic
coordinates. Only clusters in superclusters with at least 8 members are
plotted. The supergalactic $Y=0$ plane
coincides almost exactly with the galactic equatorial plane.
}
\includegraphics{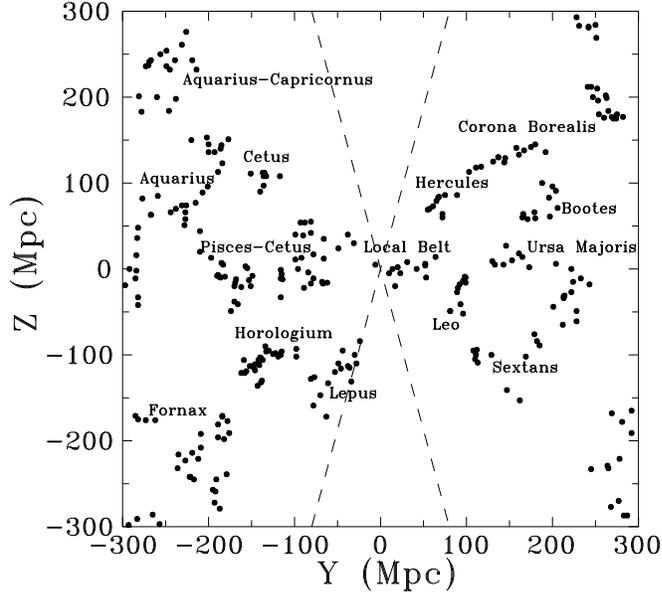}
\label{fig1}
\end{figure*}

\section{Distribution of clusters of galaxies}

The distribution of clusters of galaxies located in very rich
superclusters with at least 8 member-clusters is shown in Figure~1~
 \cite{me}. We see a rather regular network of superclusters and voids.
High-density regions are separated from each other by a rather constant
intervals of $\approx 120$~\Mpc. Such quasiregular distribution resembles
honeycombs or 3D chessboard \cite{tully}.

\begin{figure*}[t]
\vspace*{6cm}
\caption{
The power spectrum $P(k)$ for clusters with measured redshifts is plotted
with solid circles; $2\sigma$ error bars are shown.  The solid line is the
standard CDM ($h =0.5$, $\Omega = 1$) power spectrum enhanced by a bias
factor of $b=3$.}
\includegraphics{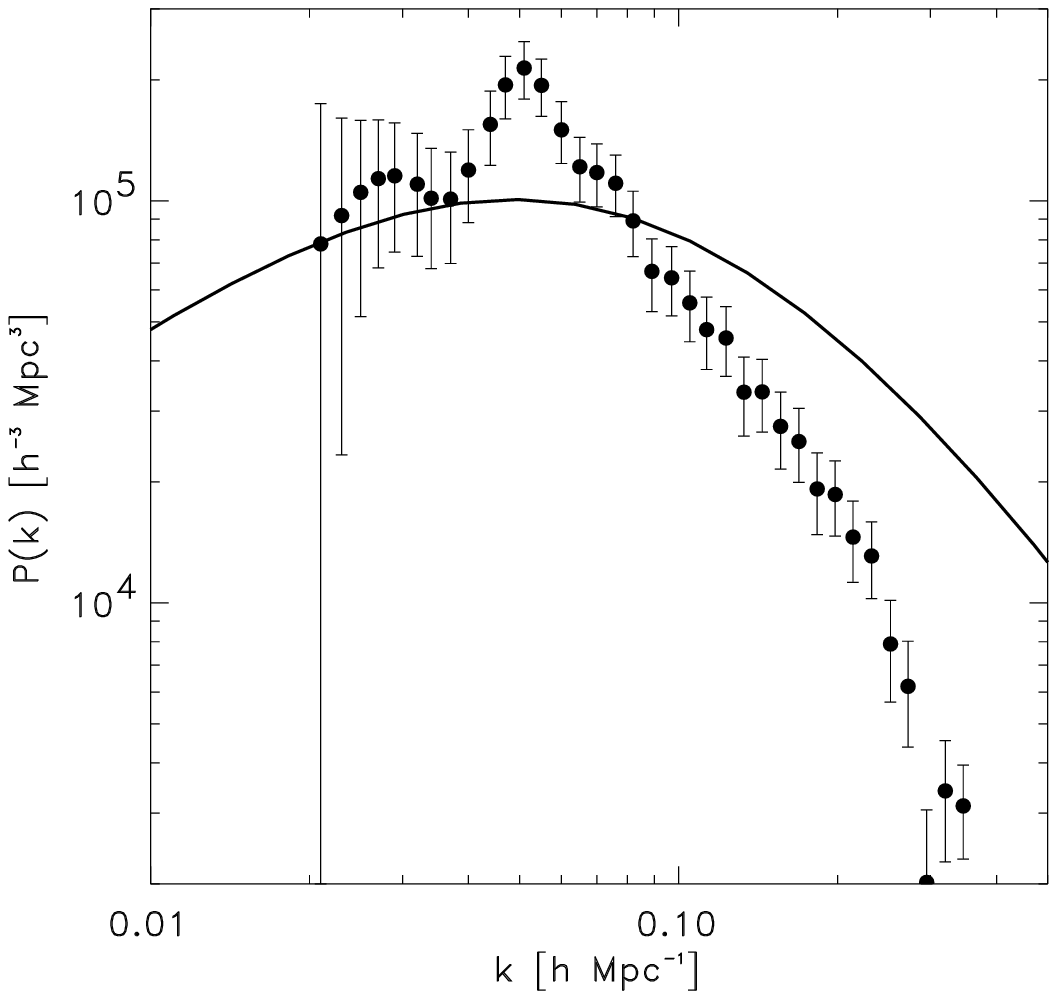}
\label{fig2}
\end{figure*}

The correlation function of quasiregularly located sample of clusters of
galaxies should have periodic maxima and minima which correspond to mutual
distances of high- and low-density regions. Such a phenomenon is actually
observed \cite{einasto2}. The power spectrum of clusters of galaxies has a
peak on wavelength $\lambda_0=120$~\Mpc\ which corresponds to the size of
the step of the supercluster-void network \cite{einasto} (see Figure~2).
On short wavelengths the spectrum  can be approximated by a power law with
index $n=-1.8$.  On long wavelengths  the spectrum is compatible with the
Harrison-Zeldovich spectrum with power index $n=1$.

This unusual form of the power spectrum raises the question: Is such a
peaked power spectrum compatible with the angular spectrum of the
temperature anisotropy of the cosmic microwave background (CMB)?  The
temperature anisotropy spectrum has a maximum around wavenumber
$l\approx 200$ due to oscillations of the hot plasma before
recombination. This range of the spectrum has recently been measured
in Saskatoon \cite{saskatoon}. We have calculated the angular spectrum
for standard CDM-type model and a model with the peaked power spectrum
 \cite{atrio}. This test shows that using an appropriate set of
cosmological parameters the peaked power spectrum is in agreement with
CMB data. However, CMB data is not sufficient to discriminate between
the standard CDM-type and the peaked power spectrum.

A density field with a smooth power spectrum as produced in CDM-type
models will generate a random distribution of high- and low-density
regions  \cite{einasto2}, in contrast to recent cluster data.  This
leads us to the conclusion that the combined evidence from cluster and
CMB data favour models with a built-in scale in the {\it initial}
spectrum.  Double inflation models provide a possible scenario where
the formation of a peak could have taken place. A version of the
double inflation model is suggested by
Starobinsky \cite{starobinski}. The study of the distribution of
matter on large scales could provide a direct test of more complicated
models of inflation.

\vskip0.2cm

I thank H. Andernach, F. Atrio-Barandela, M. Einasto, S.  Gottl\"ober,
V. M\"uller, A. Starobinsky, E. Tago and D. Tucker for fruitful
collaboration and permission to use our joint results in this talk.
This study was supported by the Estonian Science Foundation.

\end{document}